\documentclass{webofc}
\usepackage[varg]{txfonts}   % Web of Conferences font

\newcommand{\tr}{\operatorname{tr}}

\newcommand{\bbR}{\mathbb{R}}

\newcommand{\bbZ}{\mathbb{Z}}

\newcommand{\scri}{\mathcal{I}}

\begin{document}
\title{Towards causal patch physics in dS/CFT}

\author{\firstname{Yasha} \lastname{Neiman}\inst{1}\fnsep\thanks{\email{yashula@gmail.com}}}

\institute{Okinawa Institute of Science and Technology, 1919-1 Tancha, Onna-son, Okinawa 904-0495, Japan}

\abstract{
	This contribution is a status report on a research program aimed at obtaining quantum-gravitational physics inside a cosmological horizon through dS/CFT, i.e. through a holographic description at past/future infinity of de Sitter space. The program aims to bring together two main elements. The first is the observation by Anninos, Hartman and Strominger that Vasiliev's higher-spin gravity provides a working model for dS/CFT in 3+1 dimensions. The second is the proposal by Parikh, Savonije and Verlinde that dS/CFT may prove more tractable if one works in so-called ``elliptic'' de Sitter space -- a folded-in-half version of global de Sitter where antipodal points have been identified. We review some relevant progress concerning quantum field theory on elliptic de Sitter space, higher-spin gravity and its holographic duality with a free vector model.
	We present our reasons for optimism that the approach outlined here will lead to a full holographic description of quantum (higher-spin) gravity in the causal patch of a de Sitter observer.
}
\maketitle
\section{Introduction to higher-spin dS/CFT} \label{sec:intro}

For decades, the great conceptual problem in theoretical physics has been that of quantum gravity. In the real world, this problem remains unsolved -- we don't know how nature reconciles the two simple facts $\hbar\neq 0$ and $G\neq 0$. However, one can also frame quantum gravity as a less demanding question: how can these two facts be reconciled even in principle, in any fictitious but mathematically consistent universe? This version of the question is now essentially solved. The first fully convincing solution (by the author's standards) was the discovery of AdS/CFT in the late 1990's \cite{Maldacena:1997re,Witten:1998qj,Aharony:1999ti}. There, a non-gravitating conformal field theory (CFT) provides a complete description of quantum-gravitational observables at spatial infinity, in a universe with negative cosmological constant $\Lambda < 0$. Remarkably, at around the same time that AdS/CFT was discovered by theorists, astronomical observations \cite{Riess:1998cb,Perlmutter:1998np} have falsified its core premise: it appears that our Universe has  a cosmological constant of the ``wrong sign'', i.e. $\Lambda > 0$. Thus, despite some wide-spread impressions, quantum gravity is both theoretically tractable and observationally falsifiable: an entire class of theoretical models has been found, and has been falsified by observation! Naively, this is very strange: one would expect that any observation relevant to quantum gravity should involve the tiny Planck scale. Instead, due to the way in which holography interchanges UV and IR, we find a crucial dependence of the theory on $\Lambda$, i.e. on the behavior of spacetime at the \emph{largest} distances.

The sign of $\Lambda$ may at first seem like a technical detail, but in fact it has a crucial implication for our place in the Universe. In asymptotically anti-de Sitter (AdS) spacetime, with $\Lambda < 0$, \emph{spatial infinity is a physical place}. Lightlike signals can go there and back again in finite time, while massive probes and observers can get arbitrarily close. Thus, the observables calculated by AdS/CFT make operational physical sense, and it's easy to accept that only this kind of observables that can be precisely defined. After all, the bulk spacetime's geometry is experiencing quantum fluctuations, or perhaps is fundamentally ill-defined altogether. It is only at infinity, where the Planck length becomes negligible, that one can have a fixed classical geometry, and therefore pin down precise notions for what is being prepared and measured. All of this is quite satisfactory and consistent, except it is not applicable to the real world. In asymptotically de Sitter (dS) spacetime, with $\Lambda > 0$, spatial infinity is completely irrelevant to observable physics, even though it may still exist from a ``God's eye'' perspective. Instead, every observer is trapped inside a cosmological horizon of finite area. In effect, when passing from AdS to dS, the roles of space and time get interchanged. In dS, instead of taking finite time to cross an infinite distance, causal signals take infinite time to cross a finite distance -- the distance to the horizon. While AdS has a conformal boundary at spatial infinity, dS has boundaries at past and future infinity. While the AdS boundary is a physically meaningful \emph{place}, the dS boundary is instead a pair of \emph{times}. Moreover, for every observer, most of the dS boundary is inaccessible: at infinity, the horizon radius appears pointlike, so the observer can only access the initial and final endpoints of her wordline.

Thus, we now have on our hands an updated version of the problem of quantum gravity: to reconcile the \emph{three} simple facts $\hbar\neq 0$, $G\neq 0$ and $\Lambda > 0$. The implication is \emph{to understand quantum gravity in finite regions of space}, or at least in the largest region that is physically relevant -- the causal patch of an observer in asymptotically dS spacetime. Given the success of AdS/CFT, it is tempting to try a similar approach, i.e. to construct dS/CFT \cite{Strominger:2001pn} -- holography in de Sitter space. As already covered by the above discussion, the move from AdS to dS introduces a host of conceptual issues: how to treat the two boundaries at past \& future infinity? How can time emerge holographically from a spacelike boundary? Since the boundary is inaccessible to any single observer, how do we translate the CFT's output into physics inside an observer's horizon? Is the Hilbert space of an observer's causal patch finite-dimensional, as implied by Bekenstein's entropy formula? Is this Hilbert space the same for all observers? What are the implications for the measurement procedure in quantum mechanics, which relies on a measurement apparatus with an infinite-dimensional state space? 

For a number of years, these issues were widely discussed -- see e.g. \cite{Witten:2001kn,Balasubramanian:2002zh,Dyson:2002pf,Parikh:2002py,Maldacena:2002vr,Banks:2002wr,Klemm:2004mb,ArkaniHamed:2007ky}. However, active interest has gradually died down, due to the absence of a reliable working model on which to test the various ideas. Indeed, while AdS/CFT models abound, the vast majority of them turn pathological if we try to flip the sign of $\Lambda$. There are two simple reasons for this. First, string theory in general and AdS/CFT in particular are well-understood only in the presence of unbroken supersymmetry, which is incompatible with the dS spacetime symmetry structure. Second, in addition to the graviton and other massless or light fields, string theory generically contains a tower of massive excited modes. When infinity is a time instead of a place, these massive fields oscillate -- in analogy with the $e^{-imt}$ behavior in Minkowski space -- which implies complex conformal weights on the boundary. Both of these problems are difficult, though perhaps potentially solvable. For the time being, if one wants a working model of dS/CFT, one had better work with a theory that does not require supersymmetry, and does not contain any massive fields. In, 2011, a model of just this type has been put forward \cite{Anninos:2011ui}.

The model of \cite{Anninos:2011ui} is a duality between type-A higher-spin gravity in $dS_4$ and a free vector model on the 3d boundary at future infinity. It results from flipping the sign of $\Lambda$ in the corresponding AdS/CFT model of \cite{Klebanov:2002ja}. Higher-spin gravity \cite{Vasiliev:1995dn,Vasiliev:1999ba} is an interacting theory of an infinite tower of massless fields with increasing spin; in the minimal version, there is one field for every even integer spin $(s=0,2,4,\dots)$. The infinite tower of massless fields corresponds to an infinite-dimensional gauge symmetry -- higher-spin symmetry -- which extends the ordinary spacetime symmetry of General Relativity (GR). In the holographic model of \cite{Klebanov:2002ja,Anninos:2011ui}, a global version of the infinite-dimensional higher-spin symmetry is unbroken, which can be realized in the boundary CFT if and only if the latter is free \cite{Maldacena:2011jn}. One can think of higher-spin gravity as the ``opposite of supergravity'': it extends the spacetime gauge symmetry in a bosonic direction rather than a fermionic one. Unlike supergravity, though, higher-spin theory is very different from GR coupled to matter, since the spin-2 graviton is now joined by fields of spin $s>2$. Moreover, unlike in string theory, where most of the fields are heavy and decouple at low energy, here all of the higher-spin fields are massless, and they all interact at once and at all orders in derivatives. As a result, the interactions of higher-spin theory are \emph{nonlocal at the cosmological scale}, and are nothing like those of GR. Thus, this model is highly unrealistic. However, it permits us to do holography with $\Lambda >0$, and, as an added bonus, does not require the extra dimensions of string theory: the bulk is really (A)dS\textsubscript{4}, without the additional compact factor that is always present in stringy AdS/CFT.

By the time the model of \cite{Anninos:2011ui} was proposed, work on the conceptual underpinnings of dS/CFT has largely died down. The main conceptual framework that was left standing, and was used in \cite{Anninos:2011ui}, is that of \cite{Maldacena:2002vr}. There, the Lorentzian dS space is treated as a natural analytic continuation of Euclidean AdS, sharing the same symmetries and the same boundary. Using the well-understood Euclidean AdS/CFT dictionary, this leads one to recognize the CFT partition function as the \emph{Hartle-Hawking wavefunction \cite{Hartle:1983ai}} of the Lorentzian dS at future infinity. While this interpretation is very solid, it fails to address the issue of cosmological horizons: the wavefunction at future infinity is not accessible to any observer, and a dictionary is still needed between that and the physics inside an observable patch.

Our goal, then, is to extract the physics within causal patches for the higher-spin dS/CFT model of \cite{Anninos:2011ui}. In other words, we wish to revisit the old conceptual questions of dS/CFT, this time within the context of a working model. Specifically, we will pick up the suggestion of \cite{Parikh:2002py}, where it was proposed that dS/CFT should become more tractable on a folded-in-half version of de Sitter space, where every observer can access all of space (though not all of spacetime). We contend that, while this idea has some well-known problems when applied to GR, it is a match made in heaven with the higher-spin model. If our marriage between the two proves successful, we will have the first-ever working model of quantum (sort-of-)gravity inside a cosmological horizon. Given the great differences between real-world GR and higher-spin theory, such a model may not address all the outstanding questions about $\Lambda > 0$ quantum gravity. However, some questions, such as the dimension of the causal patch's Hilbert space, should be general enough that they can be usefully addressed even within this toy system. 

\section{Folding de Sitter space in half}

Global de Sitter space is the hyperboloid of constant spacelike radius in the ambient flat spacetime $\bbR^{1,4}$:
\begin{align}
 dS_4 = \left\{x^\mu\in\bbR^{1,4}\, |\, x_\mu x^\mu = 1 \right\} \ . \label{eq:dS}
\end{align}
The asymptotes to this hyperboloid are the past-pointing and future-pointing null directions in $\bbR^{1,4}$. These make up the boundaries of $dS_4$ at past and future infinity:
\begin{align}
 \scri^\pm = \left\{\ell^\mu\in\bbR^{1,4}\, |\, \ell_\mu\ell^\mu = 0 \ ; \ \ell^0 \gtrless 0 \ ; \ \ell^\mu \cong \lambda\ell^\mu \ \ \forall\lambda>0 \right\} \ . \label{eq:scri}
\end{align}
An observer in $dS_4$ is specified by her worldline's asymptotic endpoints $p_i\in\scri^-$ and $p_f\in\scri^+$. We identify the observable patch as the region inside both past and future horizons; for a detailed argument on this point, see \cite{Halpern:2015zia}.

\begin{figure}%
	\centering%
	\includegraphics[scale=0.7]{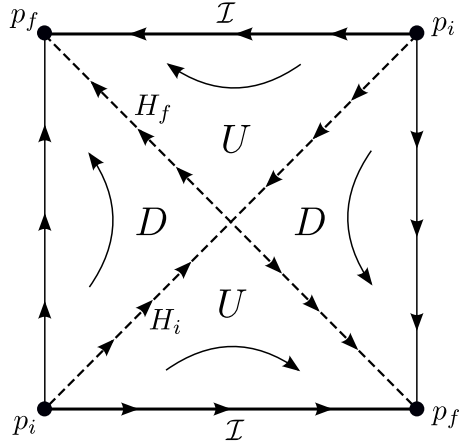} \\
	\caption{A Penrose diagram of $dS_4/\bbZ_2$. Opposite points in the diagram are identified. Past and future infinity are identified into a single boundary $\scri$. The observer's worldline endpoints $p_i,p_f\in\scri$ define horizons $H_i,H_f$, which divide the spacetime into two independent ``quadrants'': the causal patch $D$ (identified with its antipode) and its complement $U$, which contains $\scri$. The arrows denote the direction of the Killing vector $\xi^\mu$ that generates time translations in the causal patch. The observer induces a time orientation in regions where $\xi^\mu$ is causal, i.e. in the causal patch and on the horizons.}
	\label{fig:causal} 
\end{figure}

In several ways, the spacetime \eqref{eq:dS} is ``twice too big'': it has two boundaries \eqref{eq:scri} rather than one; only half the spacetime (the Poincare patch) is relevant to cosmology; and, most crucially for us, every observer sees only half of space. Thus, before we can construct a dictionary from $\scri^\pm$ into the observable patch, we must somehow disentangle the relevant information from that pertaining to the antipodal patch. 
A way out was suggested by \cite{Parikh:2002py}, in a modern echo of an old proposal by Schrodinger. The idea is to fold de Sitter space in half, in the unique way that preserves its full symmetry group $O(1,4)$. In the $\bbR^{1,4}$ language, this is accomplished by identifying every point $x^\mu$ with its antipodal point $-x^\mu$. The spacetime and its boundary thus become:
\begin{align}
 dS_4/\bbZ_2 = \left\{x^\mu\in\bbR^{1,4}\, |\, x_\mu x^\mu = 1 \ ; \ x^\mu \cong -x^\mu \right\} \ ; \quad \scri = \left\{\ell^\mu\in\bbR^{1,4}\, |\, \ell_\mu\ell^\mu = 0 \ ; \ \ell^\mu \cong \lambda\ell^\mu \ \ \forall\lambda \right\} \ .
 \label{eq:elliptic}
\end{align}
The folded-in-half spacetime $dS_4/\bbZ_2$ is known as elliptic de Sitter space; the word ``elliptic'' refers to the spacelike separation between the identified pairs of points $\pm x^\mu$. The spacetime \eqref{eq:elliptic} has only one boundary: past and future infinity are identified with each other. Crucially, every observer's causal patch now covers all of space, though not all of spacetime. See figure \ref{fig:causal} for the causal structure of $dS/\bbZ_2$ with a chosen observer.

As we discuss in the next section, $dS/\bbZ_2$ has a peculiar but consistent causal structure. The main reason why the $dS/\bbZ_2$ approach of \cite{Parikh:2002py} was abandoned is that this subtle consistency is not preserved by gravitational perturbations. However, we have already made the compromise of committing to the higher-spin model of \cite{Anninos:2011ui}, where the bulk theory is \emph{very different from GR}. As we discuss below, this may allow us to evade the conflict between the causal structure and bulk perturbations.

\section{Observer-dependent quantum mechanics}

The folding in half of de Sitter space carries one obvious complication: since we're identifying opposite points in both space and time, global time orientability is lost. In $dS/\bbZ_2$, there is no consistent distinction between past and future lightcones. Nevertheless, there is never an \emph{observable} violation of causality: there are no closed timelike curves, and every observer's causal patch is time-orientable. Since $\scri^-$ and $\scri^+$ have been identified, an observer in $dS/\bbZ_2$ is defined by a pair of points $p_i,p_f$ on the single boundary $\scri$. The time orientation in the causal patch is then induced by the ordering of $p_i$ and $p_f$, i.e. by the orientation of the observer's worldline. Different observers induce time orientations on different causal patches. Where their causal patches intersect, the time orientations of two different observers may or may not agree.

This peculiar status of the arrow of time carries crucial implications for quantum mechanics in $dS/\bbZ_2$. Indeed, the operator algebra $[q,p] = i$ for e.g. a particle depends on our convention for the sign of the velocity, and therefore on the signs of time derivatives. As a result, while quantum mechanics is consistent in observable patches of $dS/\bbZ_2$, there is no global, observer-independent quantum mechanics of the entire Universe. This situation is reminiscent of the idea of horizon complementarity \cite{Susskind:1993if,Dyson:2002pf}. Horizon complementarity sets out to solve the black hole information paradox by positing that it depends on a consistent quantum description of the entire Universe, which need not exist. If every observer can only access a partial region of spacetime due to horizons, then a consistent quantum description of every such region should suffice. Complementarity then goes on to posit that these descriptions in the different patches are in fact equivalent -- that the same Hilbert space is being encoded in different ways.

As we can see, apart from its broader significance for holography, field theory on elliptic de Sitter space forms a unique theoretical laboratory for the idea of complementarity. Indeed, complementarity is essentially ``forced'' by the two sides of every horizon being identified. In addition, at the classical level, every observer indeed sees a different encoding of the same information: while the observers see different causal patches, each of them covers a complete time slice. Finally, unlike in most discussions of complementarity, the system is well-defined and under complete control. 

In \cite{Hackl:2014txa,Halpern:2015zia}, we carried out a detailed study of quantum field theory (QFT) in $dS/\bbZ_2$ and its observer dependence. The key idea is to introduce ordinary $dS$ as a double cover of $dS/\bbZ_2$, and to use QFT on $dS$ as the observer-independent global structure. The subtlety lies in the different mappings $dS/\bbZ_2 \rightarrow dS$ that are induced by different observers. As a result, $dS/\bbZ_2$ inherits the local field operators from $dS$, but the operator algebra becomes observer-dependent. In addition, mixed states in $dS/\bbZ_2$ naturally arise from pure antipodally symmetric states in $dS$. 

Within this framework, we were able to show that observers agree on all the expectation values in the intersections of their causal patches, provided that they consistently agree (or consistently disagree) on the time orientation. However, when two observers agree on the arrow of time in one region and disagree in another, they will generally disagree on the entanglement between these two regions. In more physical terms, they will disagree on the entanglement across one of their horizons. In particular, one observer's pure state can appear as mixed to another. As far as we can tell, this picture, while peculiar, is perfectly consistent. It therefore demonstrates that horizon complementarity may have been conceived too narrowly. In the $dS/\bbZ_2$ universe, we see a \emph{generalized} version of complementarity: the quantum world-pictures of different observers are \emph{not} fully equivalent, but they are mutually consistent where it is required.

\section{Massless holography in $dS_4/\bbZ_2$}

Having understood the structure of QFT on elliptic de Sitter space, one can proceed towards the main task: reconstructing the Lorentzian quantum physics inside a causal patch from the Euclidean partition function of the boundary CFT. As we've seen in the previous section, this reconstructed quantum physics (e.g. the operator algebra) will necessarily depend on the choice of observer. In \cite{Halpern:2015zia}, we accomplished this task in the limit of non-interacting (but quantum) bulk fields. Already at this level, it is important that we're working specifically with the higher-spin model, because our construction depends crucially on the fact that the bulk fields are \emph{massless}.

The specific problem solved in \cite{Halpern:2015zia} can be phrased as follows, using a spin-0 field for simplicity. A choice of observer in $dS_4/\bbZ_2$ breaks the spacetime symmetry group $O(1,4)$ down to the subgroup $\bbR\times O(3)$ of time translations and rotations. This suggests a basis $a_{lm}(\omega),a^*_{lm}(\omega)$ for the field's modes in the causal patch, where the frequency $\omega$ and the angular momentum numbers $(l,m)$ are conserved. Since the field is free, each of these modes is a harmonic oscillator, with $\omega$ already specified as its frequency. On the other hand, a similar decomposition into modes $c_{lm}(\omega),c^*_{lm}(\omega)$ can be performed for the boundary data at past/future infinity $\scri$ (note that the generator of time translations is spacelike outside the causal patch and tangential to $\scri$). Since the causal patch covers all of space, there must be a one-to-one mapping between the oscillators $a,a^*$ and the boundary modes $c,c^*$. Using the $\bbR\times O(3)$ symmetry and the linearity of the field equations, we see that this mapping must take the form:
\begin{align}
 a_{lm}(\omega) = N_l(\omega)c_{lm}(\omega) \ , \label{eq:a_c}
\end{align}
where $N_l(\omega)$ are some c-number coefficients. Taking the oscillators $a,a^*$ to be normalized, we can read off the \emph{operator algebra and Hamiltonian} that are induced on the Euclidean boundary data at $\scri$ by our choice of observer:
\begin{align}
&\left[\hat c_{lm}(\omega), \hat c^\dagger_{l'm'}(\omega') \right] = \frac{1}{\left|N_l(\omega)\right|^2}\cdot 2\pi\delta(\omega-\omega')\delta_{ll'}\delta_{mm'} \ ; \label{eq:c_algebra} \\
&\hat H = \int_0^\infty\frac{d\omega}{2\pi}\sum_{lm} \omega \left|N_l(\omega)\right|^2 \hat c^\dagger_{lm}(\omega)\, \hat c_{lm}(\omega) \ . \label{eq:c_Hamiltonian}
\end{align}
Thus, our task is to derive the coefficients $\left|N_l(\omega)\right|^2$. Also, since we are constructing a holographic dictionary, this must be done \emph{strictly from the CFT partition function, with no bulk calculations}. This turns out to be possible, thanks to a peculiar property of \emph{free massless bulk fields}: their two types of boundary data (Dirichlet vs. Neumann, electric vs. magnetic) correspond directly to the two possible signs of antipodal symmetry, i.e. to the two possible kinds of fields in the $dS_4/\bbZ_2$ bulk \cite{Vasiliev:2012vf,Ng:2012xp,Neiman:2014npa}. We refer to \cite{Halpern:2015zia} for the full construction, simply citing here the result:
\begin{align}
 \left|N_l(\omega)\right|^2 = S_l(\omega)\cdot \frac{1}{2}\left(\coth\frac{\pi\omega}{2}\right)^{\eta\,(-1)^l} \ , \label{eq:S_to_N}
\end{align}
where $S_l(\omega)$ are the coefficients in the Gaussian approximation to the CFT partition function:
\begin{align}
 Z_\text{CFT}\big[c_{lm}(\omega),c^*_{lm}(\omega)\big] \sim \exp\left(-\int_0^\infty\frac{d\omega}{2\pi}\sum_{lm} S_l(\omega) \left|c_{lm}(\omega)\right|^2 \right) \ . \label{eq:Z}
\end{align}

The significance of the result \eqref{eq:S_to_N} is that it shows how, as a matter of principle, one can derive a quantum operator algebra and a Lorentzian time evolution out of the CFT on the Euclidean boundary. However, since we've used so extensively the special features of free massless fields on $dS_4/\bbZ_2$, one may worry that a generalization to the interacting level will prove too difficult. If there is hope, it must lie in the fact that the interactions of higher-spin gravity are themselves very special. In the remaining sections, we will discuss some recent work in higher-spin theory that gives us cause for optimism.

\section{Higher-spin gravity on a fixed geometry}

The free-field result of the previous section relied on the $\bbR\times O(3)$ symmetry of a causal patch in $dS_4/\bbZ_2$. Even putting aside worries about causality, what are we to do once the geometry is gravitationally altered and the symmetry destroyed? At first, this problem seems even worse in higher-spin gravity than it does in GR. Indeed, in its usual formulation, higher-spin theory is ``even more'' generally covariant: the metric is not just dynamical, it is now relegated to a gauge-dependent component of the infinitely larger higher-spin connection. 

In \cite{Neiman:2015wma}, we uncovered a feature of higher-spin gravity that offers a way around this problem. Through a slight (and locally equivalent) reformulation, higher-spin theory can change from a ``more-than-gravitational'' theory on a featureless manifold into a ``non-gravitational'' theory on a fixed (A)dS\textsubscript{4} geometry! Crucially, one can do this \emph{without deforming the higher-spin gauge symmetry}, i.e. without substantially complicating the field equations. The reason why this is possible is the different relationship in GR vs. higher-spin theory between diffeomorphisms and the dynamical spin-2 field. In GR, the dynamical spin-2 field is quite crucially the ``gauge field'' for local translations, i.e. for diffeomorphisms, i.e. it must be identified with the spacetime metric. In higher-spin theory, one instead has, as a component of the higher-spin gauge group, a local ``translation'' symmetry that is completely divorced from diffeomorphisms, i.e. that only acts on fields at a given point. The reason for this is that the field equations of higher-spin gravity are written in an ``unfolded'' formulation, in which both the ordinary fields and their full tower of spacetime derivatives are packaged together into a single ``master field''. Thus, a translation can be performed by accessing the fields' derivatives that are already encoded at the given point, without the need to go to an adjacent point, i.e. without a diffeomorphism. 

With the gauge symmetry thus divorced from diffeomorphisms, there is no longer any need to identify the dynamical spin-2 field as the spacetime metric. Instead, it can live, alongside the fields of all other spins, on a fixed external geometry. We must note, though, that this external geometry must then be pure (A)dS\textsubscript{4} (or a topological modification thereof, such as $dS_4/\bbZ_2$). While this result arises naturally in \cite{Neiman:2015wma} through a mechanism peculiar to higher-spin gravity, it can be understood already at the level of free massless fields: for spin 2 and higher, the very consistency of free massless field equations on a fixed background severely constrains the allowed background geometries \cite{Penrose:1987uia}.

To sum up, on one hand, higher-spin gravity is similar to GR, and therefore interesting: it contains a spin-2 massless field, and has a diff-invariant formulation. On the other hand, it is different from GR, and therefore tractable, by virtue of having a formulation on a fixed background geometry. For our present purpose, the implication is that we can continue working on a pure $dS_4/\bbZ_2$ geometry, with its causal structure and symmetries, even at the interacting level.

\section{Bulk and CFT from twistor space}

Even if we are allowed to keep the $dS_4/\bbZ_2$ geometry at the interacting level, the interactions of higher-spin gravity must still be contended with. In the present section, we outline an angle of attack on this problem. The key intuition is that in higher-spin holography with unbroken global higher-spin symmetry, there is a sense in which \emph{the free theory is enough}. To see one way in which this is true, consider the partition function, i.e. the effective bulk ``on-shell action'', as a functional of boundary data. This can be expressed equivalently as a non-linear functional of the \emph{linearized bulk solution} corresponding to this boundary data. Now, for ordinary local theories, the linearized bulk solution is a very \emph{awkward} parameterization of the boundary data, so this observation is only of philosophical interest. In higher-spin gravity, on the other hand, thanks to the unfolded formulation, the entire bulk solution is conveniently encoded within the master field \emph{at any single point}. Thus, expressing the partition function via the linearized solution becomes practical. Moreover, the answer is almost completely fixed by higher-spin symmetry, up to an overall coefficient at every order in the perturbative expansion \cite{Colombo:2012jx,Didenko:2012tv}.

In a recent work \cite{HolographicPenrose}, we were able to go further, and obtain the complete expression for the partition function in terms of the linearized bulk solution. To accomplish this, we made use of the fact that even with the bulk interactions taken into account, the boundary CFT remains free. In other words, all the effects of bulk interactions are contained within the correlators of a free boundary theory. However, this can be put to practical use only if we express the bulk and boundary theories in some common language. This is the main achievement of \cite{HolographicPenrose}.

The common language that we used to unify bulk and boundary in the higher-spin/free-CFT model is Penrose's twistor theory \cite{Penrose:1986ca,Ward:1990vs}. Twistors are spinors of the spacetime symmetry group -- in our case, $O(1,4)$. They correspond geometrically to totally-null planes (in the complexified 4d bulk) or to null geodesics (on the complexified 3d boundary). These objects are ideally suited for describing massless fields, and indeed play a central role in the formulation of higher-spin gravity. In particular, the Penrose transform relates functions in twistor space to solutions of the free massless field equations in the 4d bulk. In \cite{HolographicPenrose}, we rewrite the Penrose transform within the geometric framework of \cite{Neiman:2015wma}, i.e. in the context of higher-spin gravity on a fixed (A)dS\textsubscript{4} geometry. We then present a ``holographic dual'' of the Penrose transform, which relates the same twistor functions to the boundary CFT's sources and operators. Finally, we express the CFT partition function in the twistor language, which is directly related via the Penrose transform to the linearized bulk solution.

Without going into any technical detail, we present here some of the key formulas. The CFT action can be written as follows, with bilocal sources $\Pi(\ell',\ell)$ coupled to the bilocal single-trace operators $\phi^I(\ell)\bar\phi_I(\ell')$, where $\ell,\ell'$ are boundary points:
\begin{align}
S_{\text{CFT}}[\Pi(\ell',\ell)] = -\int d^3\ell\,\bar\phi_I\Box\phi^I - \int d^3\ell' d^3\ell\,\bar\phi_I(\ell')\Pi(\ell',\ell)\phi^I(\ell) \ .
\end{align}
The source $\Pi(\ell',\ell)$ can be packaged into a twistor function $f(Y)$ via:
\begin{align}
 F(Y) &= -\frac{1}{4\pi}\int d^3\ell\,d^3\ell'\,\Pi(\ell',\ell) \sqrt{-2\ell\cdot\ell'}\,\delta_\ell(Y)\star\delta_{\ell'}(Y) \ ,
\end{align}
where $\delta_\ell(Y)$ is a certain delta function depending on the boundary point $\ell$, and $\star$ is the star product of higher-spin algebra. The partition function can now be written in terms of $F(Y)$ as:
\begin{align}
 Z_{\text{CFT}}[F(Y)] \sim \exp\left(-\frac{N}{4}\tr_\star\ln_\star[1+F(Y)]\right) = \left(\textstyle\det_\star[1+F(Y)]\right)^{-N/4} \ , \label{eq:Z}
\end{align}
where $\tr_\star$ is the higher-spin-invariant trace operation, and $\det_\star$ is the corresponding determinant. On the bulk side, the twistor function $F(Y)$ is related to a linearized bulk solution via the Penrose transform:
\begin{align}
 F(Y) = C(x;Y)\star i\delta_x(Y) \ ,
\end{align}
where the master field $C(x;Y)$ serves as a generating function for fields and their derivatives at the bulk point $x$, while $\delta_x(Y)$ is a bulk version of $\delta_\ell(Y)$.

\section{Outlook and summary}

In this note, we motivated higher-spin holography in ``elliptic'' de Sitter space $dS_4/\bbZ_2$ as a promising approach for developing the first-ever working model of quantum (higher-spin) gravity inside a cosmological horizon. We described the successful solution \cite{Halpern:2015zia} of this problem at the level of free bulk fields, and presented some results to indicate that the full interacting theory may prove tractable as well. The next step in this program should be to rewrite the free-field result of \cite{Halpern:2015zia} in the twistor language of \cite{HolographicPenrose}, described in the previous section. Then, if we are lucky, the full interacting result will readily present itself, similarly to the partition function \eqref{eq:Z} in \cite{HolographicPenrose}.

In the work described here, we are making heavy use of the very special and wildly unrealistic properties of higher-spin gravity. As a result, even if we succeed at extracting quantum physics inside the horizon, many of the original questions of $\Lambda>0$ quantum gravity may not even make sense within our treatment. However, this shouldn't be the case for \emph{every} interesting question. In particular, we hope to find out whether the Hilbert space inside the horizon is finite-dimensional, and whether its dimension is captured by the Bekenstein-Hawking black hole entropy formula.

Finally, we must note that higher-spin gravity is of interest for reasons other than its applicability to $\Lambda>0$. The developments described in the previous two sections are of general relevance to the study of this theory. A successful formulation of the quantum physics inside a causal patch of $dS_4/\bbZ_2$ should be of similarly broad relevance, as it will constitute a more direct quantization of higher-spin gravity than what is currently available.

\section*{Acknowledgements}

I thank the organizers of the ICGAC-13/IK-15 conference at Ewha Womans University for their hospitality. Parts of the work reported here were performed in collaboration with Lucas Hackl and Illan Halpern. YN is supported by the Quantum Gravity Unit of the Okinawa Institute of Science and Technology Graduate University (OIST). During earlier stages of the work, YN was employed at Perimeter Institute and Penn State University, where he was supported by the Government of Canada through Industry Canada, by the Province of Ontario through the Ministry of Research \& Innovation, by NSERC Discovery grants, as well as by the NSF grant PHY-1205388 and the Eberly Research Funds of Penn State.


\begin{thebibliography}{}
	
\bibitem{Maldacena:1997re} 
J.~M.~Maldacena,
``The Large N limit of superconformal field theories and supergravity,''
Int.\ J.\ Theor.\ Phys.\  {\bf 38}, 1113 (1999)
[Adv.\ Theor.\ Math.\ Phys.\  {\bf 2}, 231 (1998)]
doi:10.1023/A:1026654312961
[hep-th/9711200].

\bibitem{Witten:1998qj} 
E.~Witten,
``Anti-de Sitter space and holography,''
Adv.\ Theor.\ Math.\ Phys.\  {\bf 2}, 253 (1998)
[hep-th/9802150].

\bibitem{Aharony:1999ti} 
O.~Aharony, S.~S.~Gubser, J.~M.~Maldacena, H.~Ooguri and Y.~Oz,
``Large N field theories, string theory and gravity,''
Phys.\ Rept.\  {\bf 323}, 183 (2000)
doi:10.1016/S0370-1573(99)00083-6
[hep-th/9905111].

\bibitem{Riess:1998cb} 
A.~G.~Riess {\it et al.} [Supernova Search Team],
``Observational evidence from supernovae for an accelerating universe and a cosmological constant,''
Astron.\ J.\  {\bf 116}, 1009 (1998)
doi:10.1086/300499
[astro-ph/9805201].

\bibitem{Perlmutter:1998np} 
S.~Perlmutter {\it et al.} [Supernova Cosmology Project Collaboration],
``Measurements of Omega and Lambda from 42 high redshift supernovae,''
Astrophys.\ J.\  {\bf 517}, 565 (1999)
doi:10.1086/307221
[astro-ph/9812133].

\bibitem{Strominger:2001pn} 
A.~Strominger,
``The dS / CFT correspondence,''
JHEP {\bf 0110}, 034 (2001)
doi:10.1088/1126-6708/2001/10/034
[hep-th/0106113].

\bibitem{Witten:2001kn} 
E.~Witten,
``Quantum gravity in de Sitter space,''
hep-th/0106109.

\bibitem{Balasubramanian:2002zh} 
V.~Balasubramanian, J.~de Boer and D.~Minic,
``Notes on de Sitter space and holography,''
Class.\ Quant.\ Grav.\  {\bf 19}, 5655 (2002)
[Annals Phys.\  {\bf 303}, 59 (2003)]
doi:10.1016/S0003-4916(02)00020-9
[hep-th/0207245].

\bibitem{Dyson:2002pf} 
L.~Dyson, M.~Kleban and L.~Susskind,
``Disturbing implications of a cosmological constant,''
JHEP {\bf 0210}, 011 (2002)
[hep-th/0208013].

\bibitem{Parikh:2002py} 
M.~K.~Parikh, I.~Savonije and E.~P.~Verlinde,
``Elliptic de Sitter space: dS/Z(2),''
Phys.\ Rev.\ D {\bf 67}, 064005 (2003)
[hep-th/0209120].

\bibitem{Maldacena:2002vr} 
J.~M.~Maldacena,
``Non-Gaussian features of primordial fluctuations in single field inflationary models,''
JHEP {\bf 0305}, 013 (2003)
[astro-ph/0210603].

\bibitem{Banks:2002wr} 
T.~Banks, W.~Fischler and S.~Paban,
``Recurrent nightmares? Measurement theory in de Sitter space,''
JHEP {\bf 0212}, 062 (2002)
doi:10.1088/1126-6708/2002/12/062
[hep-th/0210160].

\bibitem{Klemm:2004mb} 
D.~Klemm and L.~Vanzo,
``Aspects of quantum gravity in de Sitter spaces,''
JCAP {\bf 0411}, 006 (2004)
doi:10.1088/1475-7516/2004/11/006
[hep-th/0407255].

\bibitem{ArkaniHamed:2007ky} 
N.~Arkani-Hamed, S.~Dubovsky, A.~Nicolis, E.~Trincherini and G.~Villadoro,
``A Measure of de Sitter entropy and eternal inflation,''
JHEP {\bf 0705}, 055 (2007)
doi:10.1088/1126-6708/2007/05/055
[arXiv:0704.1814 [hep-th]].

\bibitem{Anninos:2011ui} 
D.~Anninos, T.~Hartman and A.~Strominger,
``Higher Spin Realization of the dS/CFT Correspondence,''
arXiv:1108.5735 [hep-th].

\bibitem{Klebanov:2002ja} 
I.~R.~Klebanov and A.~M.~Polyakov,
``AdS dual of the critical O(N) vector model,''
Phys.\ Lett.\ B {\bf 550}, 213 (2002)
[hep-th/0210114].

\bibitem{Vasiliev:1995dn} 
M.~A.~Vasiliev,
``Higher spin gauge theories in four-dimensions, three-dimensions, and two-dimensions,''
Int.\ J.\ Mod.\ Phys.\ D {\bf 5}, 763 (1996)
[hep-th/9611024].

\bibitem{Vasiliev:1999ba} 
M.~A.~Vasiliev,
``Higher spin gauge theories: Star product and AdS space,''
In *Shifman, M.A. (ed.): The many faces of the superworld* 533-610
[hep-th/9910096].

\bibitem{Maldacena:2011jn} 
  J.~Maldacena and A.~Zhiboedov,
  ``Constraining Conformal Field Theories with A Higher Spin Symmetry,''
  J.\ Phys.\ A {\bf 46}, 214011 (2013)
  [arXiv:1112.1016 [hep-th]].
  
\bibitem{Hartle:1983ai} 
J.~B.~Hartle and S.~W.~Hawking,
``Wave Function of the Universe,''
Phys.\ Rev.\ D {\bf 28}, 2960 (1983).

\bibitem{Halpern:2015zia} 
I.~F.~Halpern and Y.~Neiman,
``Holography and quantum states in elliptic de Sitter space,''
JHEP {\bf 1512}, 057 (2015)
doi:10.1007/JHEP12(2015)057
[arXiv:1509.05890 [hep-th]].

\bibitem{Susskind:1993if} 
L.~Susskind, L.~Thorlacius and J.~Uglum,
``The Stretched horizon and black hole complementarity,''
Phys.\ Rev.\ D {\bf 48}, 3743 (1993)
[hep-th/9306069].

\bibitem{Hackl:2014txa} 
L.~Hackl and Y.~Neiman,
``Horizon complementarity in elliptic de Sitter space,''
Phys.\ Rev.\ D {\bf 91}, no. 4, 044016 (2015)
[arXiv:1409.6753 [hep-th]].

\bibitem{Halpern:2015zia} 
I.~F.~Halpern and Y.~Neiman,
``Holography and quantum states in elliptic de Sitter space,''
JHEP {\bf 1512}, 057 (2015)
doi:10.1007/JHEP12(2015)057
[arXiv:1509.05890 [hep-th]].

\bibitem{Vasiliev:2012vf} 
M.~A.~Vasiliev,
``Holography, Unfolding and Higher-Spin Theory,''
J.\ Phys.\ A {\bf 46}, 214013 (2013)
[arXiv:1203.5554 [hep-th]].

\bibitem{Ng:2012xp} 
G.~S.~Ng and A.~Strominger,
``State/Operator Correspondence in Higher-Spin dS/CFT,''
Class.\ Quant.\ Grav.\  {\bf 30}, 104002 (2013)
[arXiv:1204.1057 [hep-th]].

\bibitem{Neiman:2014npa} 
Y.~Neiman,
``Antipodally symmetric gauge fields and higher-spin gravity in de Sitter space,''
JHEP {\bf 1410}, 153 (2014)
[arXiv:1406.3291 [hep-th]].

\bibitem{Neiman:2015wma} 
Y.~Neiman,
``Higher-spin gravity as a theory on a fixed (anti) de Sitter background,''
JHEP {\bf 1504}, 144 (2015)
doi:10.1007/JHEP04(2015)144
[arXiv:1502.06685 [hep-th]].

\bibitem{Penrose:1987uia} 
R.~Penrose and W.~Rindler,
``Spinors And Space-time. Vol. 1: Two-Spinor Calculus and Relativistic Fields,'' section 5.8,
Cambridge, Uk: Univ. Pr.

\bibitem{Colombo:2012jx} 
N.~Colombo and P.~Sundell,
``Higher Spin Gravity Amplitudes From Zero-form Charges,''
arXiv:1208.3880 [hep-th].

\bibitem{Didenko:2012tv} 
V.~E.~Didenko and E.~D.~Skvortsov,
``Exact higher-spin symmetry in CFT: all correlators in unbroken Vasiliev theory,''
JHEP {\bf 1304}, 158 (2013)
[arXiv:1210.7963 [hep-th]].

\bibitem{HolographicPenrose} 
Y.~Neiman,
``The Holographic Dual of the Penrose Transform,''
arXiv:1709.08050 [hep-th].

\bibitem{Penrose:1986ca} 
R.~Penrose and W.~Rindler,
``Spinors And Space-time. Vol. 2: Spinor And Twistor Methods In Space-time Geometry,''
Cambridge, Uk: Univ. Pr. (1986) 501p

\bibitem{Ward:1990vs} 
R.~S.~Ward and R.~O.~Wells,
``Twistor geometry and field theory,''
Cambridge, UK: Univ. Pr. (1990) 520p

\end{thebibliography}
\end{document}